\documentstyle[12pt]{article}

\catcode`\@=11
\long\def\@makefntext#1{
\protect\noindent \hbox to 3.2pt {\hskip-.9pt
$^{{\ninerm\@thefnmark}}$\hfil}#1\hfill}		

\def\@makefnmark{\hbox to 0pt{$^{\@thefnmark}$\hss}}  

\def\ps@myheadings{\let\@mkboth\@gobbletwo
\def\@oddhead{\hbox{}
\rightmark\hfil\ninerm\thepage}
\def\@oddfoot{}\def\@evenhead{\ninerm\thepage\hfil
\leftmark\hbox{}}\def\@evenfoot{}
\def\sectionmark##1{}\def\subsectionmark##1{}}

\renewcommand{\thefootnote}{\fnsymbol{footnote}}

\newcounter{sectionc}\newcounter{subsectionc}
\newcounter{subsubsectionc}
\renewcommand{\section}[1] {\vspace*{0.6cm}\addtocounter{sectionc}{1}
\setcounter{subsectionc}{0}\setcounter{subsubsectionc}{0}\noindent
	{\normalsize\bf\thesectionc. #1}\par\vspace*{0.4cm}}
\renewcommand{\subsection}[1]
{\vspace*{0.6cm}\addtocounter{subsectionc}{1}
	\setcounter{subsubsectionc}{0}\noindent
	{\normalsize\it\thesectionc.\thesubsectionc. #1}\par\vspace*{0.4cm}}
\renewcommand{\subsubsection}[1]
{\vspace*{0.6cm}\addtocounter{subsubsectionc}{1}
	\noindent {\normalsize\rm\thesectionc.
\thesubsectionc.\thesubsubsectionc.
	#1}\par\vspace*{0.4cm}}

\newcounter{appendixc}
\newcounter{subappendixc}[appendixc]
\newcounter{subsubappendixc}[subappendixc]

\renewcommand{\appendix}[1] {\vspace*{0.6cm}
        \refstepcounter{appendixc}
        \setcounter{figure}{0}
        \setcounter{table}{0}
        \setcounter{equation}{0}
        \renewcommand{\thefigure}{\Alph{appendixc}.\arabic{figure}}
        \renewcommand{\thetable}{\Alph{appendixc}.\arabic{table}}
        \renewcommand{\theappendixc}{\Alph{appendixc}}
        \renewcommand{\theequation}{\Alph{appendixc}.
\arabic{equation}}
        \noindent{\bf Appendix \theappendixc #1}\par\vspace*{0.4cm}}

\def\abstracts#1{{
	\centering{\begin{minipage}{12.2truecm}
\footnotesize\baselineskip=12pt\noindent
	\centerline{\footnotesize ABSTRACT}\vspace*{0.3cm}
	\parindent=0pt #1
	\end{minipage}}\par}}

\renewenvironment{thebibliography}[1]
	{\begin{list}{\arabic{enumi}.}
	{\usecounter{enumi}\setlength{\parsep}{0pt}
\setlength{\leftmargin 1.25cm}{\rightmargin 0pt}
	 \setlength{\itemsep}{0pt} \settowidth
	{\labelwidth}{#1.}\sloppy}}{\end{list}}

\newcounter{itemlistc}
\newcounter{romanlistc}
\newcounter{alphlistc}
\newcounter{arabiclistc}

\newcommand{\fcaption}[1]{
        \refstepcounter{figure}
        \setbox\@tempboxa =
\hbox{\footnotesize Fig.~\thefigure. #1}
        \ifdim \wd\@tempboxa > 6in
           {\begin{center}
        \parbox{6in}{\footnotesize\baselineskip=12pt
 Fig.~\thefigure. #1}
            \end{center}}
        \else
             {\begin{center}
             {\footnotesize Fig.~\thefigure. #1}
              \end{center}}
        \fi}

\newcommand{\tcaption}[1]{
        \refstepcounter{table}
        \setbox\@tempboxa = \hbox{\footnotesize Table~\thetable. #1}
        \ifdim \wd\@tempboxa > 6in
           {\begin{center}
        \parbox{6in}{\footnotesize\baselineskip=12pt
Table~\thetable. #1}
            \end{center}}
        \else
             {\begin{center}
             {\footnotesize Table~\thetable. #1}
              \end{center}}
        \fi}

\def\@citex[#1]#2{\if@filesw\immediate\write\@auxout
	{\string\citation{#2}}\fi
\def\@citea{}\@cite{\@for\@citeb:=#2\do
	{\@citea\def\@citea{,}\@ifundefined
	{b@\@citeb}{{\bf ?}\@warning
	{Citation `\@citeb' on page \thepage \space undefined}}
	{\csname b@\@citeb\endcsname}}}{#1}}

\newif\if@cghi
\def\cite{\@cghitrue\@ifnextchar [{\@tempswatrue
	\@citex}{\@tempswafalse\@citex[]}}
\def\citelow{\@cghifalse\@ifnextchar [{\@tempswatrue
	\@citex}{\@tempswafalse\@citex[]}}
\def\@cite#1#2{{$\null^{#1}$\if@tempswa\typeout
	{IJCGA warning: optional citation argument
	ignored: `#2'} \fi}}

 1
 1
 1

\font\ninerm=cmr9

\begin{document}

\centerline{\normalsize\bf MAGNETIC SUPERSYMMETRY BREAKING
\footnote{To
appear in the proceedings of the conference:
 "Topics in Quantum Field
Theory: Modern Methods in Fundamental Physics",
Maynooth, Ireland, May 1995. Ed. D. H. Tchrakian, World Scientific
1995.}
\footnote{This research was supported in part by
the EEC grants SC1-CT92-0792 and CHRX-CT93-0340.}
}
\baselineskip=15pt

\vspace*{0.5cm}
\centerline{\footnotesize C. P. BACHAS}
\centerline{\footnotesize\it Centre de Physique Th\'eorique,
Ecole Polytechnique}
\centerline{\footnotesize\it 91128 Palaiseau, France }
\centerline{\footnotesize E-mail: bachas@orphee.polytechnique.fr}
 \vspace*{0.5cm}
\abstracts{ I discuss the breaking of space-time
 supersymmetry when
magnetic-monopole fields are switched on in compact dimensions.
}

\setcounter{footnote}{0}
\renewcommand{\thefootnote}{\alph{footnote}}

\indent
\vskip 0.5cm
The breaking of space-time supersymmetry is arguably the main
stumbling block on the road to superstring
unification \cite{GSW}.
It raises a series of
unresolved mysteries, including those of {\it (i)} the
gauge hierarchy,  {\it (ii)} the vanishing
cosmological constant and
{\it (iii)} the stability of the dilaton and moduli.
The source
 of all these difficulties lies in the gravitational
sector, whose consistency requires   the presence of the
entire tower of string states. It is therefore
  important to have breaking
mechanisms that can be formulated directly at the string level.
Examples of such mechanisms include
the coordinate-dependent compactifications
\cite{SS},
 the anomalous-U(1) induced D-term \cite{Dterm},
 and magnetic fields
in compact dimensions \cite{Witten,magnetic}.
 The first two have been extensively
discussed in the literature.  Here I will say a few words
about the latter, referring the interested reader  to [5] for
more details.
\vskip 0.1cm

The simplest setting in which to discuss this mechanism is
 torroidal
compactifications of the type-I open superstring.
Consider for definiteness going  from ten down to four
dimensions on three 2-tori ${\cal T}_a$, where
{\ninerm a=(45),(67),(89)}.
Vacuum configurations of the $SO(32)$ gauge fields consist
of Wilson-line backgrounds, as well as
    constant magnetic
fields
 $F_a$  inside the Cartan subalgebra $U(1)^{16}$.
The magnetic fields  must
point in orthogonal directions in group space,
\begin{equation}
tr(F_aF_b)=0 \ \ \
 {\rm if} \ a\not= b   \ ,
\end{equation}
in order that the
three-index antisymmetric tensor ${\cal H}^{MNR}$
of string theory  be single-valued. They must also
obey the Dirac quantization conditions
\begin{equation}
 F_a = k_a Q_a/{\cal A}_a \ ,
\end{equation}
where ${\cal A}_a$ is the area of the $a$th torus,
$Q_a$ an $SO(32)$ generator normalized so that the smallest
charge in the adjoint representation equals one,
 and the $k_a$
are arbitrary integers. That the
above backgrounds solve the tree-level equations
 is a peculiarity of open-string theory,
in which gravitational
(closed-string) tadpoles are postponed to the one-loop
level \cite{Callan}.
The full  string spectrum can be  summarized by the
formula: \begin{equation}
\delta {\cal M}^2 = \sum_{a=(45),(67),(89)}
  (2 n_a + 2 \Sigma_a +1)\
\epsilon_a \ ,
\end{equation}
where the $n_a$ are integers labelling the Landau levels,
$\Sigma_a$ is the spin operator on the $a$th plane,
 and
$\epsilon_a$ is a non-linear function of the field $F_a$
and of the two charges on the string endpoints.
In the weak-field limit one has
\begin{equation}
\epsilon_a \simeq q_a k_a/{\cal A}_a \ ,
\end{equation}
with $q_a$ the total charge of the state in
 the direction $Q_a$.
We have not indicated in equation (3) the
spin-independent contributions to the mass coming from
  Wilson-line backgrounds on the tori.
 These only contribute when the
string state does not feel the corresponding magnetic
field, i.e. when
the appropriate charge $q_a=0$.
\vskip 0.1cm

The pattern of supersymmetry breaking is encoded in the
mass formula (3). Notice that only the charged states
are affected, and these belong   to $10d$ vector
multiplets, i.e. $N=4$ multiplets in four dimensions.
Such   multiplets include
 a space-time vector with internal-spin
 assignement $(\Sigma_{(45)} ,
\Sigma_{(67)} ,\Sigma_{(89)} ) = (0,0,0) $,
four chiral fermions with
 spin assignement
 $(\pm{1\over 2}, \pm{1\over 2}, \pm{1\over 2})$
where the  number of plus signs must be  even, and
  six space-time
scalars with internal-spin assignements
 $(\pm 1,0,0)$, $(0,\pm 1,0)$
and $(0,0,\pm 1)$.
The splitting of the multiplet that follows
 easily from the above
can
 exhibit
two very interesting features:
{\it chirality} and   {\it tachyonic
scalars}.
The net number of massless chiral fermions
is in fact a consequence of  the index theorem,
\begin{equation}
\# chiral - \# antichiral = \prod_a k_a q_a \ ,
\end{equation}
and does not generically vanish.
The   tachyonic scalars on the other hand   are
  a manifestation of the well-known
Nielsen-Olesen instability \cite{NO} of non-abelian magnetic
backgrounds. Their presence
  can trigger gauge-symmetry breaking, with appropriate
reduction of the  rank of the group.
One can exploit these features to find simple
compactifications
with a classical spectrum coming remarkably close to that
of the standard model \cite{magnetic}.
\vskip 0.1cm

Whether  such  features can survive after all
the dust in the gravitational sector
 settles down is unclear.
Besides gravitational instabilities, the
  magnetic breaking
  shares in fact with the other stringy mechanisms one
extra problem
\cite{decomp}:
the splittings are proportional to the
inverse size of some compact dimensions,
 so that one has a
non-renormalizable field theory between the string and
supersymmetry-breaking scales. If these scales are
hierarchically different there is a risk of
blowing-up coupling constants \cite{Kaplu}.
 One possible way out of this difficulty
 is to have a   tree-level   breaking
which is large but  confined
to a hidden  hypermassive sector. The splittings then get
  hierarchically suppressed when transferred gravitationally
to the observable sector. This is reminiscent of
gaugino-condensation scenarios \cite{gaugino}, but
could be implemented with magnetic fields
directly at the string level.
Such a solution
 cannot by the way be envisaged
 in the context of Scherk-Schwarz
compactifications which give a universal mass  to
gauginos \cite{SS}.  The difficulty with the above scenario
is, however, the large induced cosmological constant.



\end{document}